\documentclass[aps,prd,preprint,superscriptaddress,9pt]{revtex4-1}
\pdfoutput=1

\usepackage{graphicx}  
\usepackage{latexsym}
\usepackage{slashed}
\usepackage{dcolumn}   
\usepackage{bm}        
\usepackage{amsmath}
\usepackage{amssymb}   
\usepackage{longtable}

\def\beq{\begin{equation}}
\def\eeq{\end{equation}}
\def\bey{\begin{eqnarray}}
\def\eey{\end{eqnarray}}
\newcommand{\be}{\begin{equation}}
\newcommand{\ee}{\end{equation}}
\newcommand{\bea}{\begin{eqnarray}}
\newcommand{\eea}{\end{eqnarray}}
\newcommand{\bma}{\begin{matrix}}
\newcommand{\ema}{\end{matrix}}

\newcommand{\bml}{\begin{mathletters}}
\newcommand{\eml}{\end{mathletters}}
\newcommand{\bes}{\begin{subequations}}
\newcommand{\ees}{\end{subequations}}

\newcommand{\bi}{\begin{itemize}}
\newcommand{\ei}{\end{itemize}}

\newcommand{\lagrange}{\affiliation{Department of Chemistry and Physics, LaGrange College, LaGrange, GA 30240, USA}}

\begin{document}




\title{Possible Couplings of Dark Matter}

\author{Kevin J. Ludwick}
\email{kludwick@lagrange.edu} \lagrange



\begin{abstract}

Dark matter interacts gravitationally, but it presumably interacts weakly through other channels, especially with respect to regular luminous matter.  We look at different ways in which dark matter may couple to other fields.  We briefly review some example approaches in the literature for modeling the coupling between dark energy and dark matter and examine the possibility of an arguably better-motivated approach via non-minimal coupling between a scalar field and the Ricci scalar, which is necessary for renormalization of the scalar field in curved space-time.  We also show an example of a theory beyond the Standard Model in which dark matter is uniquely connected to the inflaton, and we use observational 
astrophysical constraints to specify an upper bound on the dark matter mass.  In turn, this mass constraint implies a limit on the unification scale of the theory, a decoupling scale of 
the theory, and the number of $e$-folds of inflation allowed.  
\end{abstract}
\maketitle

\renewcommand{\thepage}{\arabic{page}}
\setcounter{page}{2}
\renewcommand{\thefootnote}{\#\arabic{footnote}}


\begin{center}
{\bf I.   ~~  INTRODUCTION}
\end{center}


It is fascinating to think that only roughly 4\% of our universe is made up of ordinary matter that we are familiar with, while dark matter and dark energy comprise the rest.  We still do 
not understand the fundamental nature of dark matter or dark energy.  

Dark matter has only been detected gravitationally so far, and the candidates for dark matter include macroscopic objects, such as black holes and massive compact halo objects (MACHOs), and many non-baryonic particle models \cite{Peebles}, including weakly interacting massive particle (WIMP) models.  Dark matter was first inferred from the rotation curves of galaxies \cite{Zwicky, Rubin}, which seemed to indicate that there must be some unseen mass providing the gravitational potential needed for the orbiting rates of stellar matter near the outer reaches of galaxies to be as high as what was observed.  Direct detection experiments that look for direct interaction between dark matter and a target material have strongly constrained the allowed cross section for many interactions due to non-observation \cite{Buchmueller, Liu}, and indirect detection may potentially come from the detection of decay products \cite{Conrad, Halzen}, such as neutrinos that the IceCube experiment may detect \cite{IceCube}, or cosmic rays accelerated by supernovae that the AMS-02 experiment has studied \cite{AMS}.  There is currently a 3.5-keV radiation signature coming from certain galaxies (and which is noticeably absent in others) that may be explained by interactions with dark matter \cite{1405.7943}.  For more review of dark matter, consider [11-13].

In the following, we present interesting aspects of some possible dark matter couplings.  We examine a connection between dark matter and other fields via non-
minimal coupling (i.e., 
coupling to other fields through the Ricci scalar).  After briefly reviewing some parametrizations of coupled dark matter and dark energy in the literature, we 
explore in detail the coupling 
between dark energy and dark matter that must be present simply due to space-time curvature by making some reasonable and general assumptions about the dark energy potential and the coupling strength, and we are able to 
describe the conversion between dark energy and dark matter without ever explicitly specifying a coupling parametrization.  Next, we describe a model beyond the Standard Model called the 
luminogenesis model, which incorporates in a consistent way the inclusion of dark matter and the inflaton, along with other particles beyond the Standard Model.  We describe the unique coupling 
between dark matter and the inflaton in this model, and we use astrophysical constraints to arrive at an upper bound on the dark matter mass, which in turn constrains the unification 
scale and another scale of the luminogenesis model, along with the number of $e$-folds of cosmic inflation allowed. 
 
\bigskip

\begin{center}
{\bf II.   ~~  COUPLED DARK MATTER AND DARK ENERGY}
\end{center}

Consider the action for general relativity in which dark energy is represented by a real scalar field ($c=1$):
\begin{equation}
\label{action}
S = S_g + S_\phi + S_\xi + S_m = \int d^4 x \sqrt{-g} \left[ \frac{R}{16 \pi G} - \frac{1}{2} g^{\mu \nu} \nabla_\mu \phi \nabla_\nu \phi - V(\phi) - \frac{1}{2} \xi R \phi^2  \right] + S_m,
\end{equation}
where the first term is the usual contribution to the Einstein tensor ($S_g$), the second and third terms are the contribution to the scalar field dark energy ($S_\phi$), the fourth term allows for 
non-minimal coupling of the scalar field ($S_\xi$), and $S_m$ is the action 
for the rest of the contents of the universe.  $S_\xi$ represents the direct interaction between curvature and the scalar field, and it is necessary for the 
renormalization of a scalar field in a curved background.  Minimizing the action with respect to the metric 
leads to Einstein's equation, 
\begin{equation}
\label{Einstein}
R_{\mu \nu} - \frac{1}{2} R g_{\mu \nu} = 8\pi G T_{\mu \nu} \equiv 8\pi G(T_{\mu \nu}[\phi]+T_{\mu \nu}[m]),
\end{equation}
where 
\begin{align}
T_{\mu \nu}[m] = -\frac{2}{\sqrt{-g}} \frac{\delta S_m}{\delta g^{ \mu \nu}}, \label{stressm} \\
T_{\mu \nu}[\phi] \equiv -\frac{2}{\sqrt{-g}} \frac{\delta( S_\phi + S_\xi)}{\delta g^{ \mu \nu}}. \label{stressphi} 
\end{align}
We have included the variation of the interaction term in $T_{\mu \nu}[\phi]$.  There are different ways of accounting for $S_\xi$ \cite{0002091}.  Some choose to include the 
variation of $S_\xi$ instead in the form of an effective gravitational constant $G_{eff}$ that varies with $\phi$, but we 
choose to have a constant $G$ with an altered stress-energy tensor for $\phi$.  And it follows that 
\begin{equation}
\label{Econs}
\nabla_\mu  T^{\mu \nu} = 0.
\end{equation}  
Each component of the contents of the universe is typically 
modeled as a perfect fluid so that in the fluid's rest frame 
\begin{equation}
T_{\mu \nu}[i] = \mathrm{diag}(\rho_i,p_i,p_i,p_i), 
\end{equation}
where $i$ stands for either $\phi$ or some other content of the universe,
$\rho_i$ is its fluid energy density, and $p_i$ is its fluid pressure.  

In standard cosmology, the flat Friedmann-Lema\^{i}tre-Robertson-Walker 
(FLRW) metric, which describes a homogeneous and isotropic universe, is typically used:
\begin{equation}
\label{FLRWmetric}
ds^2 = - dt^2 +a^2(t)\left(dx^2 +dy^2 + dz^2 \right).
\end{equation}
Using this metric, the solutions to Einstein's equations are called the Friedmann equations:
\begin{align}
H^2 = \frac{8 \pi G}{3} \rho, \label{Friedmann1} \\
\dot{H} + H^2 = - \frac{4\pi G}{3} (\rho + 3p), \label{Friedmann2}
\end{align}
where $H \equiv \dot{a}/a$ and $\cdot$ represents differentiation with respect to $t$.  

Energy-momentum conservation, Eq. (\ref{Econs}), implies
\begin{equation}
\label{continuity}
\dot{\rho} + 3H (\rho+p) = 0.
\end{equation}
This equation can also be obtained from Eqs. (\ref{Friedmann1}) and (\ref{Friedmann2}) and so is not independent of these.  Minimizing the action 
with respect to the field $\phi$ results in the equation of motion 
\begin{equation}
\label{EOM}
\ddot{\phi} + 3 H \dot{\phi} + V'(\phi) + \xi R \phi = 0,
\end{equation}
where $'$ represents differentiation with respect to $\phi$.  

In the concordance model of cosmology, each component of the universe is assumed to be separately conserved, that is, 
\begin{equation}
\label{concordancecont}
\dot{\rho_i} + 3H (\rho_i+p_i) = 0
\end{equation}
for all $i$.  In an interacting fluid model, the total fluid is conserved, but not each component separately.  If we consider the late universe dominated by dark matter and dark energy, 
then 
\begin{equation}
\rho = \rho_\phi + \rho_m ~ \mathrm{and} ~ p = p_\phi + p_m, 
\end{equation}
and the interaction between the dark matter and dark energy fluids is typically described as 
\begin{align}
\dot{\rho_\phi} + 3H (\rho_\phi+p_\phi) = -Q, \label{Qphi} \\
\dot{\rho_m} + 3H (\rho_m+p_m) = Q.
\end{align}
A sampling of proposals for the interaction term $Q$ are as follows: 
\begin{align}
Q = \beta H \rho_\phi. \\ 
Q = \beta H \rho_m, \\
Q = \beta H (\rho_m + \rho_\phi), \\
Q = \beta H \rho_\phi \rho_m /(\rho_\phi + \rho_m), \\
Q = -\beta (\dot{\rho_\phi}+\dot{\rho_m}). 
\end{align}
The third interaction term listed here has been used as an approach toward solving the coincidence problem.  For more details on these models and others see the 
review \cite{1603.08299}.  It has also been shown that some amount of interaction between dark energy and dark matter 
may alleviate tension between local measurements of $H_0$ from the Hubble Space Telescope and global measurements of $H_0$ from the Planck Satellite \cite{1801.00689}.

We are still ignorant of the fundamental nature of dark matter and dark energy, so they very well may interact directly through an interaction term coupling the dark matter and dark energy fields directly, leading to a particular form of $Q$.  At the very least, these fields should interact through the graviton.  Even more so, if $\xi$ is non-zero as 
the renormalizability of a scalar field in a curved background requires, then the form of $Q$ would be according to the term in the Lagrangian $-\frac{1}{2} \xi R \phi^2$.  This term 
is a clear indication 
of interaction since 
$R$ depends on $H$ and $\dot{H}$ in the FLRW metric, and $R$ is clearly dependent on the dark matter (and dark energy) fields via the Friedmann equations, 
Eqs. (\ref{Friedmann1}) and (\ref{Friedmann2}), since $\rho$ and $p$ can be expressed in terms of the fields, as we will show.  And even present in $\sqrt{-g}$ is a dependence on 
the field content via Einstein's equation, which relates curvature to mass-energy.  The relationship here between curvature and mass-energy 
is fixed if we treat the background as fixed.  

\bigskip

\begin{center}
{\bf A.  ~~   An Approach to the Coupling Between Dark Matter and Dark Energy}
\end{center}

We now present a clever procedure of studying the coupling between dark matter and dark energy without out directly specifying a potential $V(\phi)$ for dark energy and without 
specifying a particular parametrization for $Q$.  Using Eq. (\ref{stressphi}), one obtains \cite{1511.08736}
\begin{equation}
\label{stressphiexpression}
T_{\mu \nu} [\phi] = \nabla_\mu \phi \nabla_\nu \phi - \frac{1}{2} g_{\mu \nu} \nabla^\alpha \phi \nabla_\alpha \phi - V(\phi) g_{\mu \nu} + \xi (R_{\mu \nu} - \frac{1}{2} R g_{\mu \nu})\phi^2 + \xi (g_{\mu \nu} \box \phi^2 - \nabla_\mu \nabla_\nu \phi^2).
\end{equation}
Since 
\begin{equation} 
T_{0 0}[\phi] = \rho_\phi ~ \mathrm{and} ~ T_{i i} [\phi] = p_\phi ~ \mathrm{for} ~ i=1,2, ~\mathrm{or} ~3, 
\end{equation}
we have 
\begin{equation}
\label{rho} 
\rho_\phi = \frac{1}{2} \dot{\phi}^2+V(\phi) + 6 \xi H \phi \dot{\phi} + 3 \xi H^2 \phi^2 
\end{equation}
and 
\begin{equation}
\label{p}
p_\phi = \frac{1}{2}(1-4\xi) \dot{\phi}^2 - V(\phi) + 2\xi H \phi \dot{\phi} - 2 \xi(1-6\xi) \dot{H} \phi^2 - 3\xi(1-8\xi) H^2 \phi^2 + 2\xi \phi V'(\phi).
\end{equation}
We specify the usual equation-of-state parametrization for dark energy and dark matter, 
\begin{equation} 
p_\phi = w_\phi \rho_\phi ~ \mathrm{and} ~ p_m = w_m \rho_m, 
\end{equation}
and we assume pressureless 
dark matter, 
\begin{equation}
w_m = 0.
\end{equation}
We use the methodology and results of \cite{0905.2348} in what follows.  Instead of specifying $V(\phi)$, we simply assume that it is changes slowly.  This is a good assumption at 
least around the present cosmological time, for which  $w_\phi$ seems to be fairly constant (and close to $-1$) \cite{1106.4996}.  At the very least, a slowly changing potential is 
certainly consistent with cosmological data, and this approximation serves as a way of allowing for an explicit calculation of $w_\phi$ and $\rho_\phi$ that is valid for a variety 
of choices for $V(\phi)$.  Keeping variation small may also help minimize unknown quantum gravity effects \cite{0905.2348, 10_1, 10_2}.  

So we assume slow-roll conditions:
\begin{align}
\frac{1}{V} \frac{dV}{d\phi} \ll 1, \label{slowroll1} \\
\frac{1}{V} \frac{d^2V}{d\phi^2} \ll 1. \label{slowroll2}
\end{align}
In addition, we assume 
\begin{equation}
|w_\phi + 1| \ll 1, 
\end{equation}
meaning that $w_\phi$ is very close to $-1$, which accords with cosmological data.  We also assume 
\begin{equation}
\xi <<1
\end{equation}
for simplification, and 
this assumption is inclusive of the case in which $\xi$ is $1/6$, the conformal coupling value in four dimensions.  With these approximations, an analytic expression for $w_\phi$ can be obtained:
\begin{multline}
\label{w}
1+w_\phi(a) = \frac{1}{9} \left\{ \frac{ \big[1+(\Omega_{\phi0}^{-1}-1) a^{-3} \big] (1-\Omega_{\phi0})}{1+(a^3-1) \Omega_{\phi0}} \right\}^{2- 8\xi/3} \bigg\{6 \sqrt{2} z_0 \xi B \bigg( \big[1+(\Omega_{\phi0}^{-1}-1) a^{-3} \big]^{-1}; \frac{1}{2}-\frac{4\xi}{3},-1+\frac{4\xi}{3} \bigg)  \\
+ \big[\sqrt{3} \lambda_0(1-2\xi) - 6 \sqrt{2} z_0 \xi \big] B \bigg( \big[1+(\Omega_{\phi0}^{-1}-1) a^{-3} \big]^{-1}; \frac{3}{2}-\frac{4\xi}{3},-1+\frac{4\xi}{3} \bigg) \bigg\}^2,
\end{multline}
where a $0$ subscript denotes the present time ($a_0 = 1$),  $\Omega_{\phi0}$ is the fraction of the present dark energy density $\rho_{\phi0}$ out of the present total energy density $\rho_0$,
and we have defined 
\begin{equation}
z_0 \equiv \sqrt{\frac{4 \pi G}{3}} \phi_0 ~ \mathrm{and} ~ \lambda_0 \equiv - \frac{1}{\sqrt{8 \pi G} V} \frac{dV}{d\phi}|_{\phi=\phi_0}.  
\end{equation}
According to our assumptions, we expect $\lambda_0$ to be very small, and cosmological data for $\Omega_{\phi0}$ implies that $z_0$ should be very small, so these these $\lambda_0$ and $z_0$ can be chosen appropriately.  
The function $B(u;a,b)$ used above is the incomplete beta function:
\begin{equation}
\label{incomplete}
B(u;a,b) = \int_0^u t^{a-1}(1-t)^{b-1} dt.
\end{equation}
Under the approximations, we can express $\Omega_\phi(a)$ as 
\begin{equation}
\label{Omega}
\Omega_\phi(a)  \equiv \rho_\phi / \rho = \big[1+(\Omega_{\phi0}^{-1}-1) a^{-3} \big]^{-1}.
\end{equation}
According to the definition of the incomplete beta function, in Eq. (\ref{incomplete}), $|u|$ is less than $1$, and this is true in the case of Eq. (\ref{w}) since $u$ is equal to $\Omega_\phi (a)$, which is always less than $1$.  Also, in Eq. (\ref{incomplete}), $z$ is greater than $0$, and this implies in Eq. (\ref{w}) that $\xi$ is less than $3/8$.  

In general (no approximation), because the total pressure $p$ is only due to dark energy, 
\begin{equation}
\label{eqstate}
w \equiv \frac{p}{\rho} = \frac{p_\phi}{\rho} =  w_\phi \Omega_\phi.
\end{equation}
And using Eq. (\ref{continuity}) and 
\begin{equation}
\frac{d}{dt} = a H \frac{d}{da}, 
\end{equation}
it can be shown that in general
\begin{equation}
\rho = \rho_0 \mathrm{Exp} \bigg[- \int_1^a \frac{3 (1+w)}{a'} da' \bigg].
\end{equation}

Now we have what we need to express what $Q$ would be.  Eq. (\ref{Qphi}) tells us
\begin{equation}
-Q =  a H \frac{d \rho_\phi}{da}+3H \rho_\phi(1+w_\phi),
\end{equation}
and we can express this in terms of our expressions for $w_\phi$ and $\Omega_\phi$ from Eqs. (\ref{w}) and (\ref{Omega}) using $H$ from Eq. (\ref{Friedmann1}) and $\rho_\phi$ 
from Eq. (\ref{Omega}).  

\begin{figure}[h]
\begin{center}
\fbox{\includegraphics[scale=1]{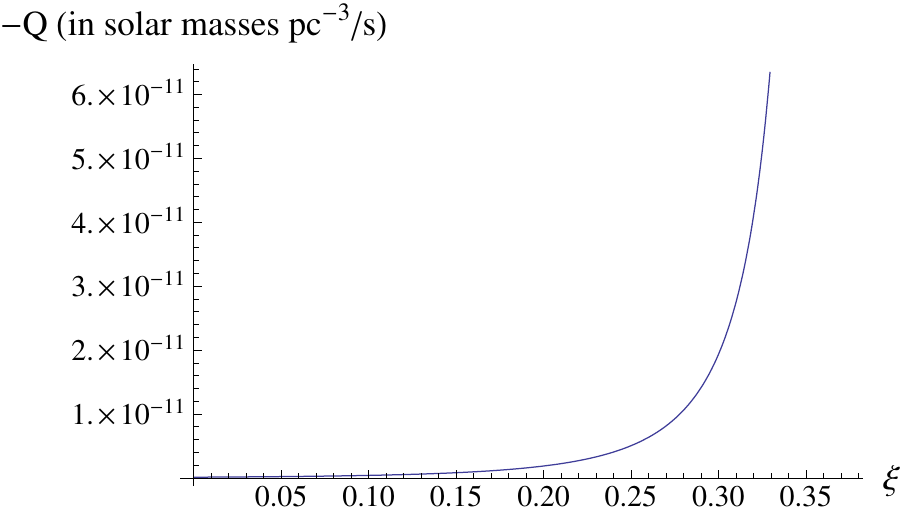}}
\caption{We plot $-Q$ (in solar masses $\times \mathrm{parsec}^{-3}$/second) vs $\xi$
for the case of redshift $z=0$, $\Omega_{\phi0} = 0.69$, $\lambda_0 = 0.01$, and $z_0 = 0.01$.  }
\label{Qxi}
\end{center}
\end{figure}

\begin{figure}[ht]
\begin{center}
\fbox{\includegraphics[scale=1]{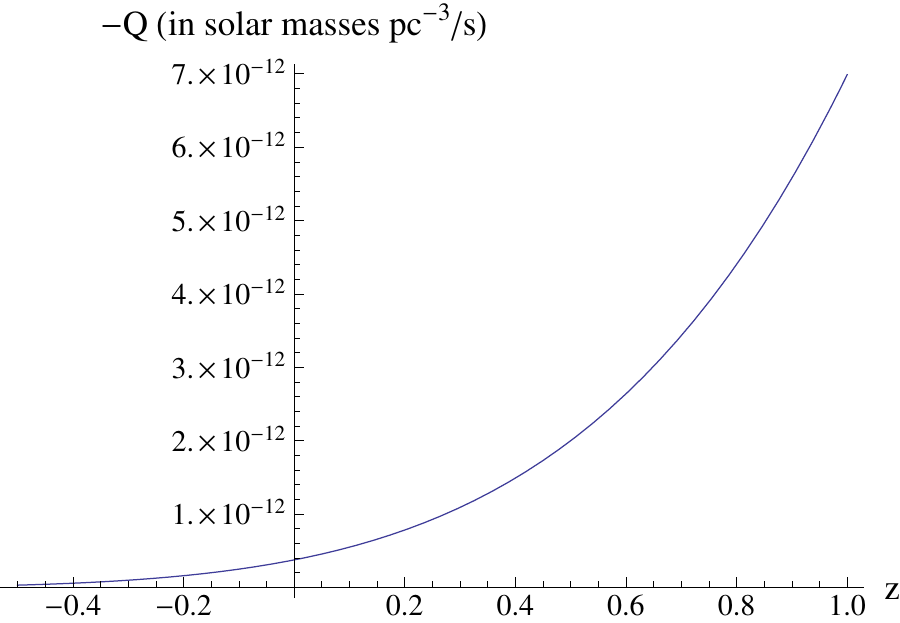}}
\caption{We plot $-Q$ (in solar masses $\times \mathrm{parsec}^{-3}$/second) vs redshift $z$
for the case of $\xi=0.1$, $\Omega_{\phi0} = 0.69$, $\lambda_0 = 0.01$, and $z_0 = 0.01$. }
\label{Qz}
\end{center}
\end{figure}

\begin{figure}[ht]
\begin{center}
\fbox{\includegraphics[scale=1]{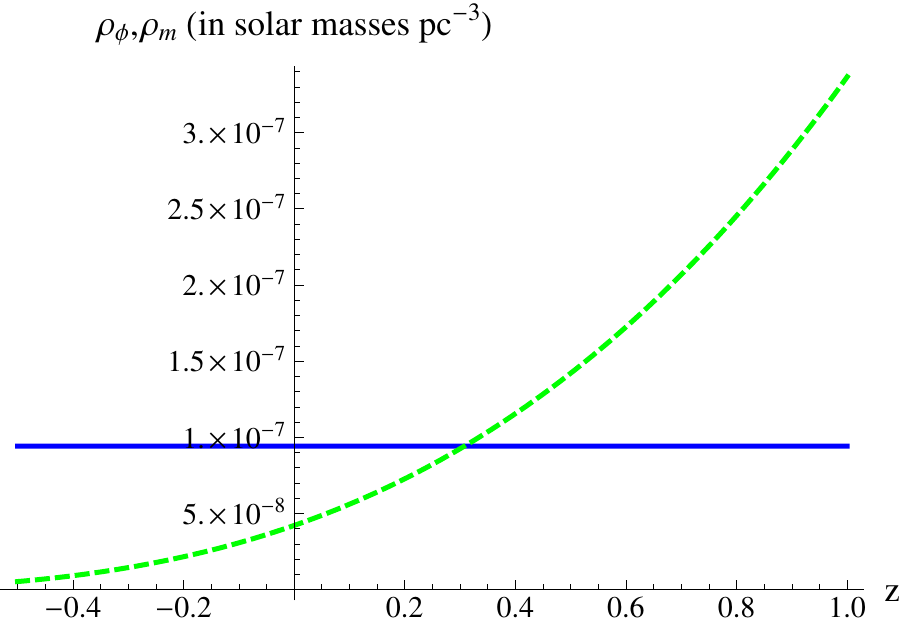}}
\caption{We plot $\rho_\phi$ and $\rho_m$ (in solar masses $\times \mathrm{parsec}^{-3}$) vs redshift $z$
for the case of $\xi=0.1$, $\Omega_{\phi0} = 0.69$, $\lambda_0 = 0.01$, and $z_0 = 0.01$. $\rho_\phi$ is represented by the blue solid line, and $\rho_m$ is represented by 
the dashed green line.}
\label{rhoz}
\end{center}
\end{figure}

As one might expect, 
for parameters that accord with cosmological data, $Q$ turns out to be very small around the present.  
In Figs. (\ref{Qxi} - \ref{rhoz}), $\Omega_{\phi0}$ is $0.69$ (in accordance with recent Planck+WP+BAO+JLA data fits from \cite{1401.4064}), and the parameters $\lambda_0$ and $z_0$ are appropriately chosen to be small: $\lambda_0 = 0.01$ and $z_0 = 0.01$.  Fig. (\ref{Qxi}) shows how $-Q$ varies with $\xi$ at the present (redshift $z=0$).The magnitude of $Q$ is small compared to the size of $\rho_\phi$ and $\rho_m$ (from Fig. (\ref{rhoz})), and we can see that the magnitude of $Q$ increases with increasing $\xi$, as one would expect 
from the $\xi$ coupling term in the Lagrangian.  Even for the case when $\xi$ is $0$, $Q$ is non-zero; although our plots here have been made using approximations, one can think of this coupling 
as due to, theoretically, the coupling of $\sqrt{-g}$ multiplying the Lagrangian in the field theory or an explicit interaction term in $V(\phi)$ that couples $\phi$ and the dark matter field directly; either way, we don't expect a large coupling.  Figs. (\ref{Qz}) and (\ref{rhoz}) show redshift $z$ on the horizontal axis ($a=\frac{1}{1+z}$), so time increases toward the left in those plots, and $z<0$ represents the future.  For both of these plots, $\xi$ is set to $0.1$.  One can 
see how $-Q$ evolves over time in Fig. (\ref{Qz}).  Fig. (\ref{rhoz}) shows how $\rho_\phi$ acts roughly as a cosmological constant (since we assumed $w_\phi \approx -1$ and strictly bigger than $-1$) and how $\rho_m$ decreases over time, as expected for cold dark matter.   

\bigskip
\begin{center}
{\bf B.  ~~   How Constraints on Dark Matter May Affect Inflation}
\end{center}

As there is currently no place for a new particle responsible for dark matter in the Standard Model of particle physics, we need a model beyond the Standard Model to include it.  
One such model is known as the luminogenesis model \cite{Luminogenesis1, Luminogenesis2, pqkevin}.  In the luminogenesis 
model, dark matter is uniquely connected to the inflaton, as we will discuss, and we are going to utilize astrophysical constraints on strongly-coupled dark matter to constrain 
its mass, which will allow us to constrain the 
unification scale and a lower scale of this theory, as well as the number of $e$-folds of inflation allowed.

The formation of galaxies and galaxy clusters is heavily influenced by the nature of dark matter.  For the usual framework of cold dark matter, there are 
discrepancies between their predictions for them and 
observations of them.  $N$-body simulations for exclusive collisionless cold dark matter predict the central density profile of dwarf galaxy and galaxy cluster halos to be very cusp-like, whereas observations indicate flat cores (cusp-vs-core problem) \cite{Blok}.  The number of Milky Way satellites predicted in simulations is bigger by an order of 
magnitude than the number inferred from observations (missing satellite problem) \cite{Moore, Klypin}, although this may not be very troublesome if more ultra-faint galaxies are successfully detected in the future \cite{Simon}.  The brightest observed dwarf spheroidal galaxy satellites of the Milky Way are predicted to be 
in the largest Milky Way subhalos, but the largest subhalos are too massive to host them (too-big-to-fail problem) \cite{Kolchin}.  The resolution of these problems may come through 
several possible means, including more 
accurate consideration of baryon interactions, astrophysical uncertainties, and warm dark matter.  A promising framework that can solve all these issues is self-interacting 
dark matter, and that is what we consider in our analysis with the luminogenesis model.  

In the luminogenesis model,  the dark and luminous sectors are unified above the Dark Unified Theory (DUT) scale.  
At this DUT scale, the unified symmetry of the model breaks ($SU(3)_C \times SU(6) \times U(1)_Y \rightarrow SU(3)_C \times SU(4)_{DM} \times SU(2)_L \times U(1)_Y  \times U(1)_{DM}$), and the breaking is triggered by the inflaton's slipping into the minimum of its symmetry-breaking (Coleman-Weinberg) potential and acquiring the true vacuum expectation value $\mu_{DUT}$, 
which is the DUT scale energy.  This symmetry breaking of $SU(6) \rightarrow  SU(4)_{DM} \times SU(2)_L \times U(1)_{DM}$ allows the 
inflaton to decay to dark matter, and dark matter can in turn decay to Standard Model (SM) and "mirror" matter.  
The representations of the luminogenesis model (which apply to each of the three families) are given below.  The existence of 
"mirror" fermions, as discussed in \cite{PQmirror1, PQmirror2}, is necessary for anomaly cancellation, and it provides a mechanism in which right-handed neutrinos may 
obtain Majorana masses proportional to the electroweak scale, and they 
could be searched for at the Large Hadron Collider.  
\begin{table}[h]
 \begin{center}
 \begin{tabular}{|l|l|lr|||} \hline
 $SU(6)$ & $SU(4)_{DM} \times SU(2)_L \times U(1)_{DM}$ \\ \hline
 ${\bf 6}$ & ${\bf (1,2)_2 + (4,1)_{-1}}$  \\
 $ {\bf 20}$ & ${\bf (4,1)_3 + (4^\ast , 1)_{-3} + (6,2)_{0} }$ \\
 ${\bf 35}$ & ${\bf (1,1)_0 + (15,1)_0 + (1,3)_0 +(4,2)_{-3}}$  \\ 
 & ${\bf  + (4^\ast , 2)_3}$ \\  \hline
 \end{tabular}
 \end{center}
 \caption{\label{table1} ${\bf (1,2)_2}$ represents luminous matter while ${\bf (4,1)_3 + (4^\ast , 1)_{-3}}$ represent dark matter.}
 \end{table}
 
The $SU(4)_{DM}$ dark matter fermions are represented by ${\bf (4,1)_3 + (4^\ast , 1)_{-3}}$ in the ${\bf 20}$ representation of $SU(6)$.  The inflaton $\phi_{inf}$ 
is represented by ${\bf (1,1)_0}$ of $\bf{35}$, and since ${\bf 20 \times 20 = 1_s+35_a+175_s+189_a}$, the inflaton decays mainly into dark matter $\chi$ through 
the interaction $g_{20} \, \Psi_{20}^{T} \sigma_2 \Psi_{20} \, \phi_{35}$, which contains the inflaton in $g_{20} \, \chi_{L}^{T} \sigma_2 \chi^{c}_{L} \phi_{inf}$.  The process of 
luminogenesis refers to the genesis of luminous matter from the initial abundance of dark matter which was formed from the decay of the inflaton.  Most indirect detectors of 
dark matter search for annihilation channels to particle-antiparticle pairs.  In the luminogenesis model, dark matter can decay to luminous particle-antiparticle pairs via an 
effective interaction with the dark photon of $U(1)_{DM}$, but also two $\chi$ particles can be converted to a fermion and mirror fermion pair.  
More details on this model can be found in the aforementioned references.  

It is assumed that ${\bf (15,1)_0 + (1,3)_0}$ ${\bf +(4,2)_{-3}+(4^{\ast},2)_3}$ of ${\bf 35}$ and ${\bf (6,2)_0}$ of ${\bf 20}$ 
have masses that are on the order of the DUT scale and thus do not affect the particle theory below that energy scale.  
Since dark matter should have no $U(1)_Y$ charge, 
the $SU(4)_{DM}$ particles in ${\bf (4,1)_{-1}}$ in the ${\bf 6}$ representation of $SU(6)$ cannot be dark matter since they have $U(1)_Y$ charge, and they are 
assumed to decouple below the mass scale we 
call $M_1$.  

In \cite{pqkevin}, we make predictions for the mass of $\chi$ in the following way:
\begin{itemize}
\item We run the $SU(2)_L$ gauge $\alpha_2$ coupling from the known electroweak scale up to some unknown DUT scale where it intersects with the $SU(4)_{DM}$ gauge coupling $\alpha_4$.
\item Then we run $\alpha_4$ down to its confinement scale, which is when $\alpha_4 \sim 1$.  In analogy with Quantum Chromodynamics (QCD) confinement of $SU(3)_C$, the main contribution to $SU(4)_{DM}$ fermions' dynamical mass is from the confinement scale of $SU(4)_{DM}$, and that energy scale is our dynamical mass prediction for $\chi$.  
\item In order to specify that scale, we need to specify a DUT scale.  Since $SU(6)$ breaks at the DUT scale when the inflaton slips into its true vacuum, we specify the DUT scale and therefore the dynamical mass of $\chi$ by constraining the parameters of a symmetry-breaking (Coleman-Weinberg) inflaton potential with Planck's constraints on the scalar spectral index and amplitude.
\end{itemize}
Using this method and the $\beta$-function equation for $SU(4)_{DM}$ and $SU(2)_L$, one can derive a formula for the dynamical dark matter mass $m_\chi$ as 
a function of the DUT scale 
energy $\mu_{DUT}$ and the scale $M_1$.  Assuming $M_1$ is the only relevant decoupling scale for $SU(4)_{DM}$ below $\mu_{DUT}$ and above the known 
electroweak scale $\mu_{EW}$, 
we have (from Eq. (10) from \cite{pqkevin})
\begin{equation}
\label{10}
m_\chi = \mathrm{Exp}\bigg[\frac{3 \pi}{19} \bigg(\frac{1}{\alpha_4(\mu_{DM})} -\frac{1}{\alpha_2(\mu_{EW})} \bigg) \bigg] M_1^{12/19} \mu_{DUT}^{8/19} ~ \mu_{EW}^{-1/19},
\end{equation}
where $\alpha_4(\mu_{DM}) \sim 1$, $\mu_{EW} = 246$GeV, and $\alpha_2(\mu_{EW}) \approx 0.03$.  We use this equation to relate $\mu_{DUT}$ to $M_1$ once we have 
obtained an upper bound on $m_\chi$ from astrophysical observational constraints.

Because of the confinement of $SU(4)$, dark baryons are formed from four $\chi$ particles.  
These particles are dubbed CHIMPs, which stands for "$\chi$ Massive Particles."  A CHIMP is denoted by $X$, and $X=(\chi \chi \chi \chi)$, and there are three dark flavors of $\chi$, 
one per luminous family of  QCD.  The three flavors enable the CHIMP to have spin zero because its wave function is a product of the $SU(4)$-color singlet wave function, which is antisymmetric, and the spin-space-flavor wave function, which can also be antisymmetric by the appropriate arrangement of 4 $\chi$s, allowing the CHIMP wave function to be symmetric. 
As we know from QCD, $SU(3)$ Nambu-Goldstone (NG) bosons appearing from the spontaneous breaking 
of chiral symmetry from $<\bar{q} q> \neq 0$ acquire a small mass from the explicit breaking of 
quark chiral symmetry due to the small masses of quarks, and they become pseudo-NG bosons known as pions.  
The small Lagrangian masses of the up and down quarks in QCD (4 and 7 MeV respectively from current algebra) in the terms $m_u \bar{u} u$ and $m_d \bar{d} d$ are much less than their dynamical masses, $\sim 300$ MeV for both, which is of the order of the QCD confinement scale $\Lambda_3$. In QCD, the so-called "constituent masses" of the up and down quarks are for the large part dynamical masses, i.e., $M_{u,d} \sim \Lambda_3$. Also, the pion mass can be obtained from the well-known Gell-Mann-Oakes-Renner relation 
\begin{equation}
m_{\pi}^2= \frac{m_u +m_d}{2} \frac{|\langle \bar{q} q\rangle|}{f_{\pi}^2}, 
\end{equation}
which shows that the pion mass vanishes as $m_u, m_d \rightarrow 0$. With $f_{\pi} \sim \Lambda_3$, it is easy to see that $m_{\pi} \ll \Lambda_3$.
Just as this results from the spontaneous 
breaking of $SU(3)_L \times SU(3)_R$ in QCD, we expect a similar phenomenon from the condensate $<\bar{\chi}_R \chi_L> \neq 0$ in 
$SU(4)$, and the NG bosons can acquire a small mass through a term $m_{0} \bar{\chi} \chi$ with $m_{0}$ a Lagrangian mass parameter for $\chi$ which, in analogy with QCD, should obey $m_{0} \ll \Lambda_4 \sim m_\chi$. Here $m_{\chi}$ is the {\em dynamical mass} which is distinct from the {\em Lagrangian mass} $m_0$. Similar to what happens in QCD, the dark pion $\pi_{DM}$ has a mass $m_{\pi_{DM}}$ proportional to $m_0$ and is expected to be small compared with the dynamical mass $m_{\chi}$. We seek to constrain the $m_{\pi_{DM}}$-$m_X$ ($m_X$ being the CHIMP mass) parameter space through astrophysical constraints via the procedure in the following section.

\bigskip

\begin{center}
{\bf C.  ~~   Solving Schr\"{o}dinger's Equation}
\end{center}

For unspecified $X$ and $\pi_{DM}$, in general, the cross section of their interaction may not lie in the regimes of the Born or classical approximations, so we cannot rely solely on analytical expressions for these regimes.  In order to find how the mass of strongly-coupled DM is correlated to the mass of a scalar mediator via astrophysical constraints, we need to numerically solve Schr\"{o}dinger's equation, 
and we use the methodology described in detail in \cite{Zurek}.  

We take the interaction between dark matter (a CHIMP, denoted by $X=(\chi \chi \chi \chi)$) and a scalar mediator ($\pi_{DM}$) to be given by an attractive Yukawa-type 
potential
\beq
V(r) = - \frac{\alpha_{DM}}{r} e^{-m_{\pi_{DM}} r} \, , \label{potential}
\eeq
where $m_{\pi_{DM}}$ is the mass parameter for $\pi_{DM}$ and the $X-\pi_{DM}$ coupling $\alpha_{DM}$ is represented by the effective interaction
\beq
\mathcal{L}_{\rm int} = g_{DM} \bar \chi \chi \pi_{DM}
\label{interaction}
\eeq
where $\alpha_{DM}$ is defined as $g_{DM}^2/(4\pi)$.  
The interaction between the CHIMPs and $\pi_{DM}$ is via the effective interaction between the scalar and the constituent 
$\chi$s in 
Eq. (\ref{interaction}), in analogy with the chiral quark model where the gluon fields have been integrated out.  Another possibility is to write an effective CHIMP-dark pion interaction Lagrangian, but then the coupling would be dimensionful.  We expect $g_{DM}$ to be at least 1 or bigger, and since the pion-nucleon coupling in QCD is $O(10)$, we analyze the cases $\alpha_{DM}=1$ and $\alpha_{DM}=10$.   

\begin{figure}[h]
\begin{center}
\fbox{\includegraphics[scale=1.4]{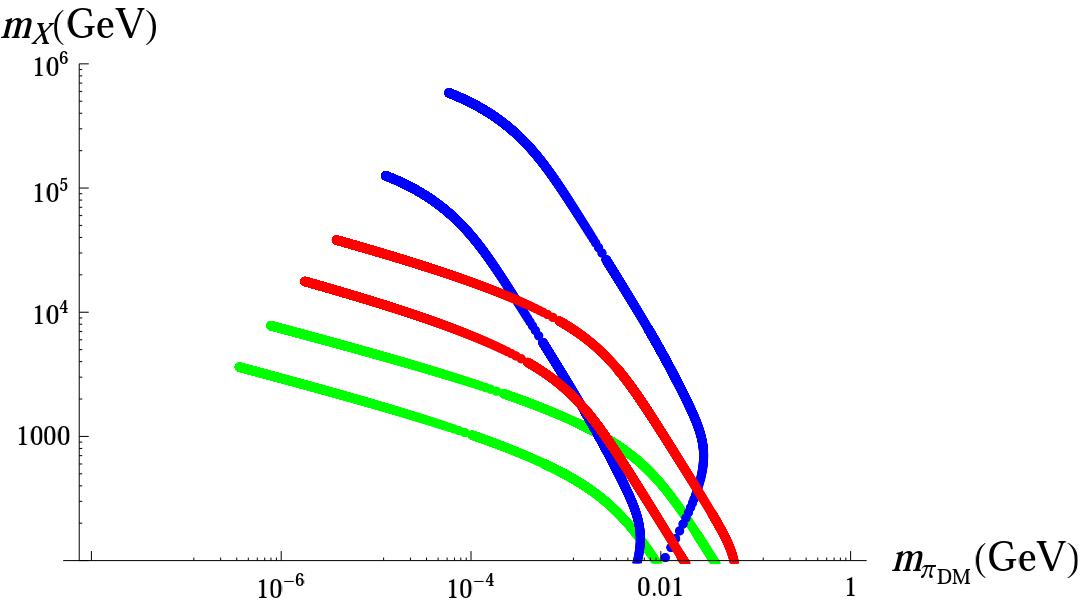}}
\caption{We plot $m_X$ vs $m_{\pi_{DM}}$ 
for the case of $\alpha_{DM}=1$.  We see that all three constraints from clusters (green), the Milky Way (red), 
and dwarf galaxies (blue) (described in the text) can be met for $m_X$ ranging from a few $100$ GeV to about $1$ TeV since this parameter space falls within all three sets of colored lines.  }
\label{1plot}
\end{center}
\end{figure}

\begin{figure}[h]
\begin{center}
\fbox{\includegraphics[scale=1]{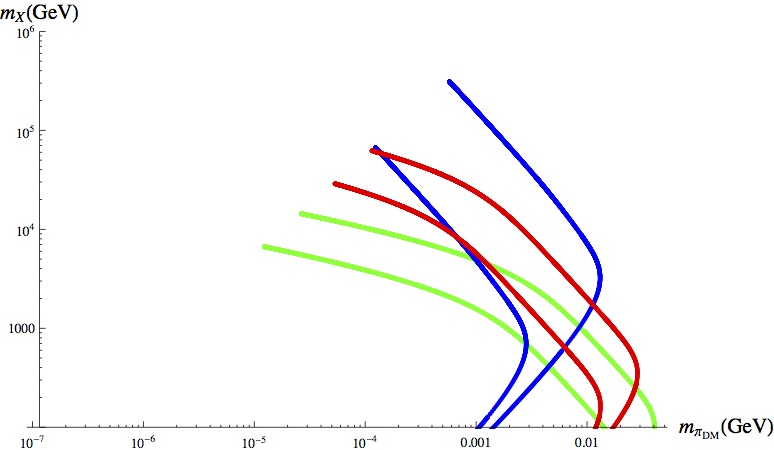}}
\caption{We plot $m_X$ vs $m_{\pi_{DM}}$ 
for the case of $\alpha_{DM}=10$.  We see that all three constraints from clusters (green), the Milky Way (red), 
and dwarf galaxies (blue) (described in the text) can be met for a range of $m_X$ with an upper limit of about $4$ TeV.  }
\label{10plot}
\end{center}
\end{figure}

We carried out the computational method for solving Schr\"{o}dinger's equation exactly as described in \cite{Zurek} with a similar level of computational accuracy for most of the steps, and we plot $m_X$ vs $m_{\pi_{DM}}$ for $\alpha_{DM}=1$ and $\alpha_{DM}=10$ via their relationship through the velocity-averaged transfer cross section $<\sigma_T>$ for the interaction described by the potential in Eq. (\ref{potential}).  The plots are shown in Figs. (\ref{1plot}) and (\ref{10plot}).

Using the convention of \cite{Zurek}, the plots are described as follows:
\begin{itemize}
\item Blue lines going from left to right respectively represent $ \langle \sigma_T \rangle/m_X =10$ and $0.1$ cm$^2$/g on dwarf scales, required for solving small scale structure anomalies.  
\item Red lines going from left to right respectively represent $\langle \sigma_T \rangle/m_X = 1$ and $0.1$ cm$^2$/g on Milky Way (MW) scales. 
\item Green lines going from left to right respectively represent $\langle \sigma_T \rangle/m_X = 1$ and $0.1$ cm$^2$/g on cluster scales. 
\end{itemize}

The above astrophysical upper and lower bounds on  $ \langle \sigma_T \rangle/m_X$ are discussed in \cite{Zurek}.  They come largely from $N$-body structure formation simulations for a limited number of specific cross sections, so their constraining 
power in our plot 
should not be taken to be extremely stringent.  But the ranges given for $ \langle \sigma_T \rangle/m_X$ are generally what is needed to satisfy observational constraints from structure formation, and we discuss the regions of $m_X$-$m_{\pi_{DM}}$ parameter space that fall within all three ranges (within the bounds of all three sets of colored lines) of $ \langle \sigma_T \rangle/m_X$.   

\bigskip

\begin{center}
{\bf D. ~~ Analysis of Results}
\end{center}

We plot the results of our analysis in Fig. (\ref{1plot}) for $m_X \geq 100$ GeV.  We are primarily interested in this mass range, and this is also the range we examined in \cite{pqkevin}.  As one can see from Fig. 6 in \cite{Zurek}, the resonances present in the three sets of constraints (blue, red, and green lines) become more aligned and overlapped as the coupling parameter $\alpha$ increases. 
We focused our computing power on calculating data points for $m_X \geq 100$ GeV since we were looking for an upper bound of mass beyond which the three sets of lines do not overlap (i.e., where all three observational constraints are not met).  
For $1 \lesssim \alpha_{DM} \lesssim 10$, we can see from Figs. (\ref{1plot}) and (\ref{10plot}) that all constraints from clusters, the Milky Way, and dwarf galaxies can be met for $m_X$ ranging from a few $100$ GeV (lower bound from the $\alpha_{DM}=1$ plot) to about $4$ TeV (upper bound from the $\alpha_{DM}=10$ plot), and this range corresponds to $1$ MeV $\lesssim m_{\pi_{DM}} \lesssim 10$ MeV.  
We point out the noteworthy observation that $m_X \gtrsim 10$ TeV does not agree with all three 
constraints in the plots (barring the fact that the tightness of these astrophysical constraints is open to interpretation, as discussed in the previous section).  

Given the numerical results in the previous paragraph, and since $\Lambda_4 \sim m_\chi \leq m_X/4$, one can see from the plots that the approximation $m_{\pi_{DM}} \ll \Lambda_4$ seems to be a good one, much 
better than the analogous chiral approximation in QCD.  This connection between the constraints on the macroscopic 
astrophysical scale and the microscopic $\pi_{DM}$-$X$ interaction lends support to the viability of the luminogenesis model.  

\begin{figure}
\begin{center}
\fbox{\includegraphics[scale=1.4]{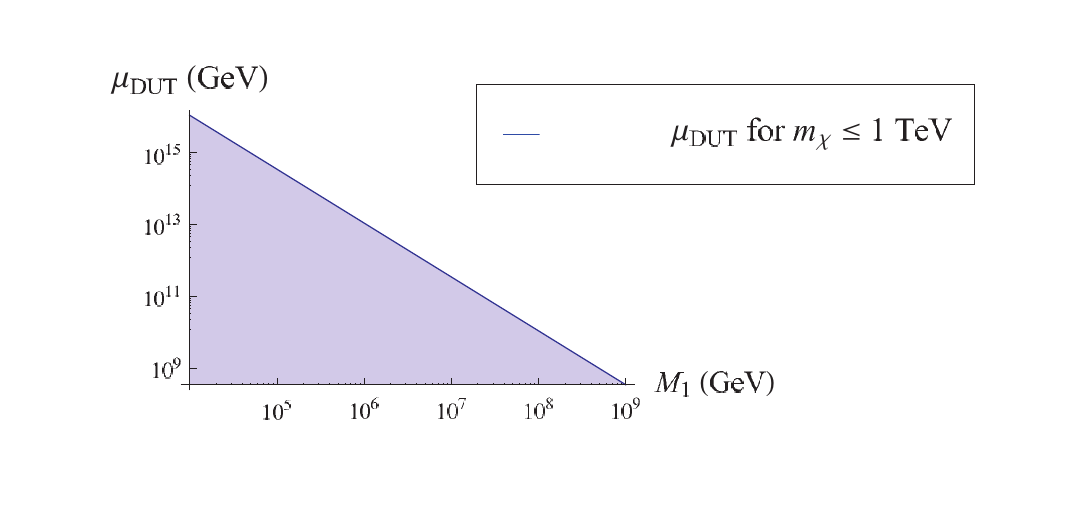}}
\caption{We plot $\mu_{DUT}$ vs $M_1$ for $m_\chi \leq 1$ TeV.  }
\label{DUTastroDM}
\end{center}
\end{figure}

We now consider the implications of this upper bound on the mass of strongly-coupled dark matter for the luminogenesis model.  Since we saw that $X=(\chi \chi \chi \chi)$ cannot have a mass bigger than about $4$ TeV, and since $m_\chi \leq m_X/4$, we 
see there is an upper bound of about $1$ TeV for $m_\chi$.  In Fig. (\ref{DUTastroDM}), we plot $\mu_{DUT}$ vs $M_1$ for this constraint $m_\chi \leq 1$ TeV using Eq. (\ref{10}).   
From Fig. (\ref{DUTastroDM}), we see that $\mu_{DUT} \leq 10^{16}$ GeV in order for this astrophysical upper bound 
for $m_\chi$ to be 
satisfied, and most of the viable parameter space (the shaded triangle) 
is for values of $\mu_{DUT}$ much less than $10^{16}$ GeV.  Along with this constraint, we also see that $M_1 \leq 10^9$ GeV to allow for $M_1 \leq \mu_{DUT}$.  

Using this upper bound on $\mu_{DUT}$ along with Planck's constraints on the scalar spectral index and amplitude, we can also determine upper bounds on the number 
of $e$-folds and the parameters of 
the potential for inflation (in our case, the Coleman-Weinberg potential we used in \cite{pqkevin}).  We work out the relationships of these parameters under the constraints 
from Planck in Eq. 21 of \cite{pqkevin}, and one can see that the number of $e$-folds would need to be less than roughly $95$.  

\bigskip
\begin{center}
{\bf IV. ~~  CONCLUSION} 
\end{center}

In general, dark matter is weakly coupled to standard luminous matter (except for gravitational coupling on large scales).  However, it is unknown how exactly dark matter interacts with non-standard entities, such as dark energy and the inflaton.  We have examined two cases of dark matter coupling.  

In the first case, we studied the coupling of dark matter to dark energy without assuming a particular functional form for the conversion rate, and we assumed that dark matter 
and dark energy were the only components present in the universe.  We illustrated a useful way of having 
interaction between dark matter and dark energy that avoided the need to specify a parametrization for $Q$, and this is convenient since we do not know what $Q$ should be from 
first principles.  We accomplished our goal by assuming a slowly varying dark energy field and a value of $\xi$ that is very small.  We pointed out that, at the very least, there should be coupling between dark matter and dark energy via the $\xi$ term in the Lagrangian necessary for the 
renormalization of the scalar field for dark energy in a curved background, and we showed in our plots that the magnitude of the coupling $Q$ indeed grew as the coupling 
constant $\xi$ increased.  
Of course, one may consider the case of scalar field dark matter, and then another term coupling this dark matter field to $R$ would be present and would indirectly represent another 
coupling of dark matter and dark energy.  Ideally, what is needed is a direct calculation of the cross section between dark energy and dark matter in curved space-time in order to 
see fundamentally how this non-minimal coupling term affects their interaction.  Also, a more accurate treatment would allow for other components of the universe to be present, 
which would allow for coupling between dark matter and regular luminous matter strictly through curvature via the Ricci scalar, although we would also expect this interaction 
to be small in general.  A more accurate treatment would also allow for back-reaction on the metric and a quantum treatment of gravity itself.     

In the second case of dark matter coupling, we showed one way that dark matter may be coupled to the inflaton.  We showed an interesting connection between the two fields in the 
luminogenesis model, which is a unified field theory that consistently combines the Standard Model with other groups that contain dark matter, the inflaton, and other non-standard 
fields.  Using constraints from $N$-body structure formation simulations, we constrained the mass of self-interacting dark matter, which in turn constrained the DUT scale and the 
$M_1$ scale of the luminogenesis model.  This constraint on the DUT scale then provided an upper limit on the number of $e$-folds of inflation allowed in the model.  

There are many potential ways in which dark matter couples to other fields, and we simply pointed out interesting facets of two different possible couplings.  The true nature of dark 
matter and how it interacts with other matter is yet to be fully unraveled, but we must pursue every feasible avenue in order to be ready when more precise measurements are 
available.  

\bigskip
\bigskip


\bigskip



\newpage

\bigskip


\begin{thebibliography}{100}

\bibitem{Peebles}
P. J. E. Peebles, {\it Nat. Astron.} {\bf 1}, 0057 (2017).
\bibitem{Zwicky}
Fritz Zwicky, {\it Helv. Phys. Acta} {\bf 6}, 110 (1933).
\bibitem{Rubin}
V. C. Rubin, {\it Publ. Astron. Soc. Pac.} {\bf 112}, 747 (2000).
\bibitem{Buchmueller}
Oliver Buchmueller, Caterina Doglioni, and Lian-Tao Wang, {\it Nat. Phys.} {\bf 13}, 217 (2017).
\bibitem{Liu}
Jianglai Liu, Xun Chen, and Xiangdong Ji, {\it Nat. Phys.} {\bf 13}, 212 (2017).
\bibitem{Conrad}
Jan Conrad and Olaf Reimer, {\it Nat. Phys.} {\bf 13}, 224 (2017).
\bibitem{Halzen}
Francis Halzen, {\it Nat. Phys.} {\bf 13}, 232 (2017).
\bibitem{IceCube}
IceCube Collaboration: M. G. Aartsen {\it et al}, {\it Phys. Rev. Lett.} {\bf 120}, 071801 (2018).  
{\tt arXiv:1707.07081 [hep-ex]}.
\bibitem{AMS}
AMS Collaboration:  M. Aguilar {\it et al}, {\it Phys. Rev. Lett.} {\bf 117}, 091103 (2016).
\bibitem{1405.7943}
Signe Riemer-S{\o}rensen, {\it Astron. Astrophys.} {\bf 590}, A71 (2016).
{\tt arXiv:1405.7943 [astro-ph.CO]}.
\bibitem{1605.04909}
Gianfranco Bertone and Dan Hooper.  
{\tt arXiv:1605.04909 [astro-ph.CO]}.
\bibitem{1201.3942}
Annika H. G. Peter.  
{\tt arXiv:1201.3942 [astro-ph.CO]}. 
\bibitem{Roberts}
M. S. Roberts, {\it ASP Conf. Series 3} {\bf 95}, 283 (2008).
\bibitem{0002091}
Valerio Faraoni, {\it Phys. Rev. D} {\bf 62}, 023504 (2000).  
{\tt arXiv:gr-qc/0002091}. 
\bibitem{1603.08299}
B. Wang, E. Abdalla, F. Atrio-Barandela, and D. Pavon, {\it Rept. Prog. Phys.} {\bf 79},  no.9, 096901 (2016).
{\tt arXiv:1603.08299[astro-ph.CO]}.
\bibitem{1801.00689}
Yang-Jie Yan, Wang Deng, and Xin-He Meng.  
{\tt arXiv:1801.00689[astro-ph.CO]}.
\bibitem{1511.08736}
Orest Hrycyna, {\it Phys. Lett. B} {\bf 768}, 218-227 (2017).
{\tt arXiv:1511.08736[astro-ph.CO]}.
\bibitem{0905.2348}
Gaveshna Gupta, Emmanuel N. Saridakis, and Anjan A. Sen, {\it Phys. Rev. D} {\bf 79}, 123013 (2009).
{\tt arXiv:0905.2348[astro-ph.CO]}.
\bibitem{1106.4996}
Paul H. Frampton, Kevin J. Ludwick, and Robert J. Scherrer, {\it Phys. Rev. D} {\bf 84}, 063003 (2011).
{\tt arXiv:1106.4996[astro-ph.CO]}.
\bibitem{10_1}
Qing-Guo Huang, {\it Phys. Rev. D} {\bf 77}, 103518 (2008).
{\tt arXiv:0708.2760[astro-ph]}.
\bibitem{10_2}
Emmanuel N. Saridakis, {\it Phys. Lett. B} {\bf 676}, 7 (2009).
{\tt arXiv:0811.1333[hep-th]}.
\bibitem{1401.4064}
SDSS Collaboration (M. Betoule {\it et al}.), {\it Astron. Astrophys.} {\bf 568}, A22 (2014).
{\tt arXiv:1401.4064[astro-ph.CO]}.
\bibitem{Luminogenesis1}
Paul H. Frampton and Pham Q. Hung, {\it Phys. Lett. B} {\bf 675}, 411 (2009). 
{\tt arXiv:0903.0358[hep-ph]}.
\bibitem{Luminogenesis2}
Paul H. Frampton and Pham Q. Hung. 
{\tt arXiv:1309.1723[hep-ph]}.
\bibitem{pqkevin}
Pham Q. Hung and Kevin J. Ludwick,  JCAP {\bf 09}, 031 (2015).
{\tt arXiv:1411.1731[hep-ph]}.
\bibitem{Blok}
W. G. J. de Blok, {\it Adv. Astron.} {\bf 2010}, 789293 (2010).  
{\tt arXiv:0910.3538[astro-ph.CO]}.
\bibitem{Moore}
Ben Moore {\it et al.}, {\it Astrophys. J.} {\bf 524}, L19 (1999).  
{\tt arXiv:astro-ph/9907411}.
\bibitem{Klypin}
Anatoly A. Klypin {\it et al.}, {\it Astrophys. J.} {\bf 522}, 82 (1999).  
{\tt arXiv:astro-ph/9901240}.
\bibitem{Simon}
Joshua D. Simon and Marla Geha, {\it Astrophys. J.} {\bf 670}, 313 (2007).  
{\tt arXiv:0706.0516[astro-ph]}.
\bibitem{Kolchin}
Michael Boylan-Kolchin, James S. Bullock, and Manoj Kaplinghat, {\it Mon. Not. Roy. Astron. Soc.} {\bf 415}, L40 (2011).  
{\tt 1103.0007[astro-ph.CO]}.
\bibitem{PQmirror1}
P.Q. Hung, {\it Phys. Lett. B} {\bf 649}, 275 (2007).
{\tt arXiv:hep-ph/0612004}. 
\bibitem{PQmirror2}
Vinh Hoang, P.Q., Hung, and Ajinkya Kamat, {\it Nucl. Phys. B} {\bf 10}, 002 (2013).
{\tt arXiv:1303.0428[hep-ph]}.
\bibitem{Zurek}
Sean Tulin, Hai-Bo Yu, and Kathryn M. Zurek, {\it Phys. Rev. D} {\bf 87}, 115007 (2013).
{\tt arXiv:1302.3898[hep-ph]}.

\end{thebibliography}
\end{document}